\documentstyle[preprint,revtex]{aps}

\newcommand{\bphi}{{\bar\phi}}

\newcommand{\bq}{\begin{eqnarray}}
\newcommand{\eq}{\end{eqnarray}}
\newcommand{\be}{\begin{equation}}
\newcommand{\ee}{\end{equation}}
\newcommand{\ba}{\begin{eqnarray}}

\newcommand{\ea}{\end{eqnarray}}
\newcommand{\baa}{\begin{eqnarray*}}
\newcommand{\eaa}{\end{eqnarray*}}
\newcommand{\barr}{\begin{array}}
\newcommand{\earr}{\end{array}}
\newcommand{\bb}{\begin{thebibliography}}
\newcommand{\eb}{\end{thebibliography}}

\newcommand{\lab}[1]{\label{#1}}

\newcommand{\Tr}{\mbox{Tr\,}}

\newcommand{\de}{\delta}
\newcommand{\ad}{\dagger}

\def\vsmall{\@setsize\vsmall{10pt}\vspt\@vspt}

\begin{document}
\draft


\begin{title}
{{\bf SYMMETRY AND OBSERVABLES IN INDUCED QCD }}
\end{title}

\vskip .3in

\author{\bf  Gordon W. Semenoff and Nathan Weiss}

\vskip 15pt

\begin{instit}
Department of Physics, University of British Columbia,\\ Vancouver,
British Columba, Canada V6T 1Z1\\ ~~\\~~\\~~\\Proceedings of \\
Mathematical Physics, String Theory and Quantum Gravity\\ Rhakov,
Ukraine, October, 1992
\end{instit}

\begin{abstract}

\centerline{\bf Abstract}

We review  some of the basic features of the Kazakov-Migdal
model of induced QCD.  We emphasize the role of $Z_N$ symmetry in
determining the observable properties of the model and also argue that
it can be broken explicitly without ruining the solvability of induced
QCD in the infinite $N$ limit.  We outline the sort of critical
behavior which the master field must have in order that the model is
still solvable.  We also review some aspects of the $D=1$ version of
the model where the partition function can be obtained analytically.
\vskip 0.5truein
\hrule

\end{abstract}

\section{\bf INDUCING QCD}

Quantum chromodynamics is presently accepted as the only viable theory
of the strong interactions.  It describes many of the quantitative
features of the interactions of hadrons at high energies.  However, at
energies lower than the hadronic scale it gives only qualitative
information.  Part of the reason is that perturbative QCD has only one
dimensionless constant, the gauge coupling which is changed
by renormalization
from a dimenisonless constant into a mass scale.  The
perturbative regime is where all external momenta of Feynman diagrams
are greater than this mass scale. It is thus impossible to address the
infrared structure of QCD, such as details of its spectrum and low
energy interactions using conventional perturbation theory.

Another parameter of QCD which could be varied is the number of colors
of quarks.  QCD is known to simplify somewhat in the limit where the
number of colors, $N$, is large \cite{lgn}.  In this limit, only
planar graphs contribute to scattering amplitudes which consequently
exhibit some of the qualitative features of the strong interactions.
However, so far no explicit solution of QCD in dimensions greater than
2 is available in the large $N$ limit and it has thus led to very few
quantitative results.

There are some features of the large N limit which are particularly
appealing.  First, there is every indication that the large $N$ limit
is quite smooth and that it exhibits confinement and dynamical chiral
symmetry breaking which is
found in the actual QCD for N=3.  Also, $1/N^2=1/9$ is
an expansion parameter which is less than one, so some large N results
should have reasonable accuracy.  Furthermore, the parameter 1/N is
not dependent on the mass scale, so once a solution is found it is
good at both high and low energies.  This might also shed some light on the
relationship between lattice and continuum QCD.

Recently, Kazakov and Migdal \cite{kmm} have proposed a novel approach
to the large N limit which has the hope of providing an exact solution
of infinite $N$ QCD.  They consider induced QCD which is obtained by
integrating over the scalar fields in the lattice gauge theory with
the partition function
\equation
Z_{\rm KM}=\int d\phi [dU]\exp\left(-N\sum_x{\rm
Tr}V[\phi(x)]+N\sum_{<x,y>} {\rm Tr}\phi(x)U(xy)\phi(y)
U^{\dagger}(xy)\right)
\label{km}
\endequation
where $\phi(x)$ are N$\times$N Hermitean matrices which reside on
lattice sites x, $U(xy)$ are unitary N$\times$N matrices which reside
on links $<xy>$ between neighboring sites x and y, $d\phi$ is the
Euclidean integration measure for Hermitean matrices, $[dU]$ is the
invariant Haar measure for integration over the unitary group U(N) and
$V[\phi]$ is a potential for the scalars.  This model is invariant
under the gauge transformations
\begin{eqnarray}
\phi(x)\rightarrow
\omega(x)\phi(x)\omega^{\dagger}(x)\\
 U(xy)\rightarrow
\omega(x)U(xy)\omega^{\dagger}(y)
\end{eqnarray}
where $\omega(x)$ is an element of U(N).  By restricting the trace of
$\phi$ to zero and the determinant of $U$ to one in (\ref{km}) we
could also consider a model with SU(N) gauge symmetry.

The partition function of the Kazakov-Migdal model (\ref{km}) can be
regarded as the $1/g^2\rightarrow\infty$ limit of lattice scalar QCD
with action
$$
	Z=\int d\phi [dU]\exp\left(-N\sum_x{\rm Tr}V[\phi(x)]+N\sum_{<x,y>}
	{\rm Tr}\phi(x)U(xy)\phi(y) U^{\dagger}(xy)+\right.
$$
\equation
	\left.+\frac{N}{g^2}\sum_{\Box}({\rm Tr}U(\Box)+{\rm
	Tr}U^{\dagger}(\Box))\right)
	\label{qcd}
\endequation
This action differs from that of the Kazakov-Migdal model by the
addition of the Wilson term,
\equation
\sum_{\Box}\left({\rm Tr}U(\Box)+{\rm
Tr}U^{\dagger}(\Box)\right)
\endequation
The Wilson term is the trace of a product of link operators around an
elementary plaquette of the lattice.  This term is the naive latticization of
the continuum Yang-Mills action, ${\rm Tr} F_{\mu\nu}F_{\mu\nu}$.  In
(\ref{qcd}) $\Box$ denotes plaquettes of the lattice and $U(\Box)$ a
product of the link operators on the links on the boundary of $\Box$.

Asymptotic freedom implies that the continuum limit of the lattice
theory (\ref{qcd}) is otained by taking the bare coupling constant to zero,
\equation
\frac{1}{g^2}\rightarrow \infty
\endequation
In fact, if instead of a lattice cutoff we had a large momentum
cutoff $\Lambda$ the bare coupling  which would be necessary
to insure one loop renormalizability of QCD is
\equation
\frac{1}{g^2}=\frac{11}{48\pi^2}\ln(\Lambda^2/\mu^2)
\endequation
The hypothesis of the Kazakov-Migdal model is that the scalar QCD
might still find a way to arrange things so that a continuum limit
exists in the opposite limit, where the bare coupling constant is
infinite
\equation
\frac{1}{g^2}\rightarrow 0
\endequation
in \cite{kmm} they present a naive argument to show how this might be
possible.  They begin with QCD coupled to scalars and without a
kinetic term for the gluon field. The Yang-Mills action is induced by
the vacuum polarization of the scalar fields in the cutoff theory.
The one-loop result is
\equation
\frac{1}{g_{\rm ind}^2}= \frac{1}{96\pi^2}\ln(\Lambda^2/m^2)
\label{bare}
\endequation
where $m$ is the scalar mass.  This can produce the $g^2$ in
(\ref{bare}) necessary to obtain a continuum limit for the gauge field
sector of the theory with the ultraviolet cutoff replaced by the
scalar mass if we take the mass of the scalar to be
\equation
m^2=\mu^2\left(\frac{\Lambda^2}{\mu^2}\right)^{1/21}
\label{beta}
\endequation
In this way, by giving up on finiteness of the scalar mass we can, at
least at one-loop order, induce a renormalizable action for QCD.  Of
course this is only a rough argument.  Higher order corrections from
hard gluons will change this result significantly.  They can only be
compensated by some strong self-interactions of the scalar field.  The
resulting picture is one of a complicated, strongly interacting
theory.  It also requires that we have the ability to arrange that the
scalar mass goes to infinity with a slower exponent than the cutoff in
the continuum limit. This is possible if the scalar field theory has a
socond order phase transition and the accompanying critical behavior.
The appealing feature of this model is that one may be able to solve
it in the large $N$ limit.

Recently we have proposed a slight modification of this idea
\cite{dksw}.  The solvability of the large $N$ limit comes about
through the absence of a kinetic term for the gauge fields in the bare
Lagrangian.  To leading order in $N$ this property is also there if
the kinetic term is not zero but is sub-leading in large $N$, or
\equation
\frac{1}{g^2}= \frac{\lambda}{N}
\endequation
where $\lambda\sim 1$ and $1/g^2\sim1/N$ as $N\rightarrow\infty$ In
\cite{dksw} we argued that, by tuning $\lambda$ appropriately we could
still produce QCD with a string tension which is finite in the
continuum limit.

The key to the solvability of the Kazakov-Migdal model is the fact
that the single--link Itzykson--Zuber integral can be done
analytically
\cite{itzub,harish}
\equation
	I_{IZ}=\int [dU] e^{N\sum \phi_i\chi_j\vert U_{ij}\vert^2}=
{\det_{(ij)}e^{\phi_i\chi_j}\over\Delta[\phi]\Delta[\chi]}
\label{itzub}
\endequation
where
\equation
\Delta[\phi]=\det_{ij}(\phi_i)^{j-1}=\prod_{i<j}(\phi_i-\phi_j)
\endequation
is the Vandermonde determinant for $\phi$.

This allows us to express the partition function (\ref{km}) as an
integral over the eigenvalues of the scalar field $\phi$ at each site
\equation
Z_{\rm KM}\propto\int\prod_{x,i} d\phi_i(x)\Delta^2[\phi(x)]
e^{-N\sum_x V[\phi_i(x)]}\prod_{<xy>} {\det_{ij}
e^{N\phi_i(x)\phi_j(y)}\over\Delta[\phi(x)]\Delta[\phi(y)]}
\label{phii}
\endequation

Here, the integral over $U$ matrices can be obtained explicitly only
because the Wilson kinetic term is absent from the action.  If the
action has a Wilson term with coefficient $\lambda$ which is of order
one in the infinite $N$ limit, the Wilson term is of order $N$ whereas
all other terms in the action are of order $N^2$.  Then, the Wilson
term can be ignored and the effective action for the scalar field, to
leading order in $N$ is still given by (\ref{phii}).

The eigenvalues $\phi_i$ behave like a master field since the large
$N$ limit in (\ref{phii}) is the classical limit and the integral can
be performed by saddle point approximation.  Migdal \cite{mig} has
derived integral equations which are obeyed by the eigenvalue density
and has given an expression for the asymptotics of the solution.
Corrections to the classical behavior and the spectrum of elementary
excitations can also be computed \cite{mig2}. (This is of course
neglecting the corrections which would arise from the presence of the
Wilson term, which should begin to contribute at this order.)  This
model has been considered further in \cite{ksw} - \cite{boulatov2}.

If the model (\ref{km}) has a second order phase transition and if the
fluctuations in the vicinity of the critical point are non-Gaussian,
one might expect that the critical behavior should be represented by
QCD, the only known nontrivial four dimensional field theory with
non-Abelian gauge symmetry.  Since the rough argument leading to
(\ref{beta}) indicates that the scalar mass should scale to infinity
slower than the lattice scale in the continuum limit, it is necessary
that the scalar field exhibits critical behavior, i.e. that the
effective scalar field theory in (\ref{phii}) has a second order phase
transition itself.

In order to familiarize the reader with the Kazakov-Migdal model, we
shall begin by reviewing how it can be solved in the simple case of a
lattice with a single site and subsequently the case where the lattice
is one-dimensional and periodic.  These simple models are interesting
in that one can obtain critical behavior for the scalar fields when
the scalar field action is quadratic.

\section{\bf KAZAKOV-MIGDAL MODEL ON A SINGLE SITE}

It is instructive to consider the Kazakov--Migdal model on a single
site.  To this end consider the following integral
\equation
	Z~=~\int {\cal D}\phi{\cal D} U ~e^{-m^2\Tr (\phi^2) +\Tr(\phi
U\phi U^{-1})}
\endequation
The evaluation of this integral follows very closely the method of
D'Adda et. al. \cite{CAP92}.  The integral over $U$ can be done using
the Itzykson--Zuber formula \cite{harish},\cite{itzub} and the result depends
only on
the eigenvalues of $\phi$.  We thus have that
\equation
	Z~\propto~\int\prod_i
d\phi_i~\Delta^2(\phi)~e^{-m^2\sum_i\phi_i^2}~ 	{{{\rm
det}\left(e^{\phi_i\phi_j}\right)}\over{\Delta^2(\phi)}}
\endequation
The determinant can be written explicitly as a sum over permutations
\equation
	Z~\propto~\sum_{\sigma\in S_N}{\epsilon(\sigma)}~
\int\prod_i d\phi_i~e^{-m^2\sum_i\phi_i^2}~
e^{\sum_i\phi_i\phi_{\sigma(i)}}
\endequation
where $\epsilon(\sigma)$ is the sign of the permutation $\sigma$.
This integral is a Gaussian integral which can be done explicitly.
The integrand can be simplified by introducing two real quantities $a$
and $b$ such that
\equation
	a^2+b^2=m^2~~~~and ~~~~ab=1
\endequation
so that
\equation
	Z~\propto~\sum_{\sigma\in S_N}{\epsilon(\sigma)}~
\int\prod_i d\phi_i~e^{-{1\over 2}\left(a\phi_i+b\phi_{\sigma(i)}
\right)^2} \label{abc}
\endequation
and
\equation
	a,b~=~m^2\pm\sqrt{m^4-1}
\endequation
Note that Eq. (\ref{abc}) is symmetric under the interchange of $a$ and $b$.

The calculation proceeds by a change of variables from the eigenvalues
$\phi_i$ to
\equation
	\xi_i=a\phi_i+b\phi_{\sigma(i)} \label{change}
\endequation
The transformation is linear and the Jacobian in independent of
$\phi$.  To evaluate the Jacobian first note that any permutation
$\sigma$ can be written as a product of r-cycles of the form
\equation
	c_1,c_2 ... c_r~\rightarrow~ c_2,c_3 ... c_r,c_1
\endequation
Consider a fixed permutation $\sigma$ and write $\sigma$ as a product
of $n_1$ cycles of size 1, $n_2$ cycles of size 2, ... $n_k$ cycles of
size $k$, ... . Clearly
\equation
	\sum_{j=1}^N jn_j~=~N
\endequation
It is also straightforward to prove by recursion that the Jacobian for
an r-cycles is
\equation
	{\rm det}~\left(\matrix {a& -b& 0& ...&\cr 0&a&-b&0...&\cr
\cr & ...&0&a&-b\cr -b&0&...&0&a\cr} 	\right)~=~a^r-b^r
\endequation
where the above is an $r\times r$ matrix.

Using the change of variables (\ref{change}) in Eq. (\ref{abc}) we find that
\equation
Z~\propto~\sum_{\sigma\in S_N}{\epsilon(\sigma)}~Z_{\sigma}
\label{sumsigma}
\endequation
with
\equation
	Z_\sigma~=~\int\prod_i d\phi_i~e^{-\xi_i^2/2} 	~=~\int
\prod_{\rm cycles}~\prod_{i\in{\rm cycles}} 	d\xi_i
{1\over{a^r-b^r}}e^{-\xi_i^2/2}
\endequation
Performing the Gaussian integral we see that
\equation
	Z_\sigma~=~\pi^{N/2}\prod_{i=1}^N\left(
{1\over{a^i-b^i}}\right)^n_i ~\propto~ 	\prod_{i=1}^\infty \left(
{1\over{a^i-b^i}}\right)^n_i
\endequation
since $n_k$$=$$0$ for $k$$>$$N$.

The next step is to do the sum over permutations (\ref{sumsigma}).  To
this end note that there are
\equation
	N!~\prod_{j=1}^N {1\over{n_j!}}{1\over{j^{n_j}}}
\endequation
distinct permutations which can be described the a given set
$(n_1,...n_N,...)$ (with $n_k$$=$$0$ for $k$$>$$N$). Thus
\equation
	Z~\propto~ \sum_{n_1=0}^\infty~\cdot\cdot\cdot
\sum_{n_k=0}^\infty \cdot\cdot\cdot 	\delta(N-\sum_{k=1}^\infty
kn_k)~ 	\prod_{j=1}^\infty \left({1\over{a^j-b^j}}\right)^{n_j}~
(-1)^{(j+1)n_j}~{1\over{k^{n_k}}}~{1\over{n_k!}}
\endequation
where we have used the fact that
\equation
	\epsilon(\sigma)=\prod_{k=1}^\infty 	(-1)^{(k+1)n_k}
\endequation
Using the integral representation of the $\delta$ function $\delta(x)
=\int_0^{2\pi} d\theta \exp (-ix\theta)$ we find that
\equation
	Z~\propto~ 	\int d\theta~e^{-iN\theta}\prod_{k=1}^\infty
\left(\sum_{n_k=0}^\infty{{ e^{i\theta kn_k}~(-1)^{(k+1)n_k}}\over
{\left[k(a^k-b^k)\right]^{n_k}~n_k!}}\right) \label{wow}
\endequation
The sum in Eq. (\ref{wow}) can be summed to an exponential. The
remaining product then becomes a sum over the exponents so that
\equation
	Z~\propto~\int d\theta
{}~e^{-iN\theta}~\exp\left[\sum_{k=1}^\infty {{(-ae^{i\theta})^k}\over
k}{1\over{1-a^{2k}}}\right]
\endequation
where we use have used the fact that $b$$=$$1/a$.  Without loss of
generality we assume that $a$$<$$1$ (i.e. $a$$=$$m^2-\sqrt {m^4-1}$.
The term $1/(1-a^{2k})$ can now be expanded in a geometric series and
the resulting sum over $k$ sums to a logarithm.  Thus
\equation
	Z~\propto~ \int d\theta~e^{-iN\theta}~\prod_{n=0}^\infty
\exp\left[\log\left(1+e^{i\theta}~a^{2n+1}\right)\right]
{}~=~\int d\theta~e^{-iN\theta}~\prod_{n=0}^\infty
\left(1+e^{i\theta}~a^{2n+1}\right)
\endequation

The final step is to perform the integration over $\theta$. To do this
note that only terms in the product which are proportional to $\exp
(iN\theta)$ will give a nonzero integral. Thus
\equation
	Z~\propto~ a^N \sum_{n_1=0}^\infty \sum_{n_2=n_1+1}^\infty
\cdot\cdot\cdot \sum_{n_N=n_{N+1}}^\infty
\left(a^2\right)^{n_1+...n_N}
\endequation
The sums can be done one at a time. They are all geometric series.
The result is
\equation
	Z~\propto~{1\over{a^N}}\prod_{k=1}^N{{\left(a^2\right)^k}\over
{1-\left(a^2\right)^k}}~=~\left(a^2\right)^{N^2/2}
\prod_{k=1}^N{1\over{{1-\left(a^2\right)^k}}}
\endequation
In fact if we define $q$$=$$a^2$$=(m^2-\sqrt{m^4-1})^2$ then
\equation
	Z~\propto~q^{N^2/2}~\left(\prod_{k=1}^N{1\over{1-q^k}}\right)
\endequation
Notice that apart from the factor $q^{N^2/2}$ this is just the
expression for a q-factorial. It is interesting to
note that a similar expression is obtained for the partition function
of a system bosons on a circle at nonzero temperature. In fact the
partition function for a bosonic string on a circle (suitably
restricted to the singlet sector) is precisely given by the above
formula.

\section{\bf KAZAKOV-MIGDAL MODEL ON A CIRCLE}

It is straightforward to generalize the above calculation to the
evaluation of the partition function of the Kazakov--Migdal Model on a
circle. This calculation is discussed in Ref. \cite{CAP92} and
\cite{kmsw1}. We consider the partition function
\equation
	Z~=~\int \prod_{x=1}^L{\cal D}\phi_x 	\prod_{x=1}^L{\cal D}
U_{x,x+1} ~e^{-m^2\sum_i\Tr (\phi_x^2) + 	\sum_x\Tr(\phi_x
U_{x,x+1}\phi_{x+1} U_{x,x+1}^{-1})}\label{onedz}
\endequation
where the fields $\phi_x$ live on the sites $x$ of a circle with
periodic boundary conditions $\phi_{L+1}=\phi_1$ and the $U_{x,x+1}$
live on the links of the circle. There are two distinct ways of
calculating this partition function.  The first method is to eliminate
almost all the $U_{x,x+1}$'s by a gauge transformation leaving only
one $U$ which cannot be eliminated. This can be chosen to be the $U$
on the first link i.e. $U$$=$$U_{1,2}$. The result is a partition
function which involves an integral over only a single link variable,
and a Gaussian integral over all $N$ Hermetian matrices.  The second
method is to integrate over all the link variables $U_{x,x+1}$
explicitly using the same formula as for the single link case above.
This second method generalizes the method presented in the previous
section for the single link integral. The details of this calculation
are discussed by D'Adda and Panzeri \cite{CAP92}.  The basic idea is to
perform all the $U$ integrations in Eq. (\ref{onedz}). The result is
\equation
	Z~\propto~\int\prod_{x=1}^L\left(\prod_{i=1}^N~d\phi_x^i\right)
\Delta^2(\phi_x)~ 	\left(\prod_{x=1}^L{{{\rm
det}~e^{\phi_x^i\phi_{x+1}^j}}\over
\Delta(\phi_x)\Delta(\phi_{x+1})}\right)
\endequation
where $\phi_x^i$ is the $i$'th eigenvalue of the matrix $\phi_x$.
Note that the Vandermonde determinants $\Delta(\phi)$ precisely cancel
leaving us with
\equation
	Z~\propto~\int \prod_{x=1}^L \left( {\rm
det}~e^{\phi_x^i\phi_{x+1}^j}
{}~e^{-m^2\sum_{i=1}^N\left(\phi_x^i\right)^2}\right)
\endequation
This is now a Gaussian integral of precisely the same form as the integral
for the K-M model on a point.  It is evaluated in Ref. \cite{CAP92}.  The
result is
\equation
	Z~\propto~q^{LN^2/2}~\left(\prod_{k=1}^N{1\over{1-q^{Lk}}}\right)
\endequation
with
\equation
	q=m^2-\sqrt{m^4-1}
\endequation
Note that this is obtained from the single link  integral by a simple
replacement
of $q$ by $q^L$.

\section{\bf SYMMETRY AND OBSERVABLES}

It was pointed out in \cite{ksw} that, like all adjoint lattice
models, the Kazakov-Migdal model (\ref{km}) has an extra gauge
symmetry which is not a symmetry of continuum QCD.  The action in
(\ref{km}) is invariant under redefining any of the gauge matrices by
an element of the center of the gauge group,
\begin{eqnarray}
U(xy) \rightarrow z(xy)U(xy)\\
\phi(x)\rightarrow\phi(x)
\label{zn}
\end{eqnarray}
where $z(xy) \in U(1)$ if the gauge group is $U(N)$ and $z(xy)\in Z_N$
if the gauge group is SU(N). (We shall call the symmetry a $Z_N$ gauge
symmetry in either case.)  It was subsequently pointed out by Gross
\cite{Gr92} and by Boulatov \cite{boulatov} that there is a larger
symmetry of this kind: one could redefine $U$ and $U^{\dagger}$ by any
element which commutes with the matrix $\phi$.  The $Z_N$ symmetry in
(\ref{zn}) is the maximal subgroup of the transformations discussed by
Gross and Boulatov which can be implemented with field independent
elements $z(xy)$. Because of this symmetry the conventional Wilson
loop observables of lattice gauge theory have vanishing average unless
they have either equal numbers of $U$ and $U^{\dagger}$ operators on
each link or else, in the case of SU(N), unless they have an integer
multiple of $N$ $U$'s or $N$ $U^{\dagger}$'s.

In conventional QCD, the expectation value of the Wilson loop operator
gives the free energy for a process which creates a heavy
quark-antiquark pair, separates them for some time and lets them
annihilate.  From the asymptotics for large loops, one extracts the
interaction potential for the quarks.  If the expectation value of the
Wilson loop behaves asymptotically like $e^{-\alpha A}$ where A is the
area of a minimal surface whose boundary is the loop, the
quark-antiquark potential grows linearly with separation at large
distances and quarks are confined. The parameter $\alpha$ is the
string tension. On the other hand if the expectation value of the
Wilson loop goes like the exponential its perimeter then the potential
is not confining.

In the Kazakov-Migdal model (\ref{km}), due to the $Z_N$ symmetry, the
expectation value of the Wilson Loop is identically zero for all loops
with non-zero area.  We can interpret this as giving an area law with
infinite string tension, $\alpha=\infty$, and no propagation of
colored objects is allowed at all. (An exception is the baryon ($U^N$)
loops in the case of SU(N) where the correct statement is that
$N$-ality cannot propagate.)  It is for this reason that the original
Kazakov-Migdal model has difficulty describing pure gluo-dynamics.

The $Z_N$ symmetry of the pure Kazakov-Migdal model is broken
explicitly by the introduction of a Wilson term in (\ref{qcd}).
However, if the Wilson term has vanishingly small coefficient and is
negligible in the large $N$ limit, one might expect the problem of
$Z_N$ symmetry to remain - the string tension would still be infinite.
It has been argued in \cite{dksw} that this need not be so.  It was
shown that, if the scalar fields exhibit a particular kind of critical
behavior, it is possible that the presence of an infinitesimally small
Wilson term is sufficient to give a finite string tension.

There are currently several other points of view on how to avoid the
constraints of $Z_N$ symmetry.  In \cite{ksw} it was suggested that if
there is a phase transition so that the $Z_N$ symmetry is represented
in a Higgs phase, the resulting large distance theory would resemble
conventional QCD.  This approach has been pursued in
\cite{KhM92},\cite{M92},\cite{KM92}.  An alternative, which was
advocated in \cite{ksw,kmsw1}, is to use unconventional observables
such as filled Wilson loops which reduce to the usual Wilson loop in
the naive continuum limit but which are invariant under $Z_N$.  The
third possibility is to break the $Z_N$ symmetry explicitly.  This was
suggested by Migdal \cite{mixed} in his mixed model in which he breaks
the $Z_N$ symmetry by introducing into the model heavy quarks in the
fundamental representation of the gauge group.

In \cite{dksw} we considered explicit $Z_N$ symmetry breaking using a
Wilson term.  We showed that it is equivalent to using the filled
Wilson loop observables.  We used that fact that the filled Wilson
loops arise naturally from ordinary Wilson loops in a modified version
of the Kazakov-Migdal model which has additional explicit symmetry
breaking terms.  We argued that one version of this modified model
should be solvable in the large $N$ limit.

\section{\bf FILLED WILSON LOOPS}

We begin with a brief review of the properties of the filled Wilson
loop operators which were introduced in \cite{ksw} and discussed in
detail in \cite{kmsw1}.  These are a special class of correlation
functions which survive the $Z_N$ symmetry of the original
Kazakov-Migdal model (\ref{km}). They are defined by considering an
oriented closed curve $\Gamma$ made of links of the lattice.  The
ordinary Wilson loop operator on $\Gamma$ is given by
\equation
	W[\Gamma]={\rm 	Tr}\left\{\prod_{<xy>\in\Gamma}U(xy)\right\}.
\endequation
For any surface $S$ which is made of plaquettes such that the boundary
of $S$ is the curve $\Gamma$ we define
\equation
	W_F[\Gamma,S]= 	W[\Gamma]\prod_{\Box\in S}W^{\dagger}[\Box]
\endequation
where $\Box$ denotes an elementary plaquette in the surface $S$. The
filled Wilson loop for $\Gamma$ is now defined as
\equation
	W_F[\Gamma]=\sum_S\mu(S)W_F[\Gamma,S]
\endequation
where the sum is over all surfaces $S$ whose boundary is the loop
$\Gamma$ with some (yet to be specified) weight function $\mu(S)$.
Notice that for each plaquette $\Box\in S$ we have inserted the
negatively oriented Wilson loop $W^{\dagger}[\Box]$.  Thus for
arbitrary weight functional $\mu(S)$ the filled Wilson loop operator
is invariant under the local $Z_N$ gauge symmetry since it has equal
numbers of U and U$^{\dagger}$ operators on each link.  Although we
have assumed that the loop is filled with {\it elementary} plaquettes
this can be easily generalized to other fillings (the other extreme
case being the adjoint loop $W[\Gamma]W^{\dagger}[\Gamma]$). We can
also define the `filled correlator'' of more than one loop by summing
over all surfaces whose boundary is given by those loops.

In ref. \cite{kmsw1} it was shown that computing the expectation value
of $W_F[\Gamma]$ is equivalent to computing the partition function of
a certain statistical model on a random two--dimensional lattice.
When computing $Z_N$ gauge invariant correlation functions of
$U$--matrices in the master field approximation the $\phi$--integral
is evaluated by substituting the master field $\bar\phi={\rm diag}
(\bar\phi_1,\ldots,\bar\phi_N)$ for the eigenvalues of $\phi$.

\equation
<U_{i_1j_1}\dots U^{\dagger}_{k_1l_1}\dots>~= 	{\int
d\phi[dU]e^{-{\rm Tr}(\sum V[\phi]-\sum \phi U\phi U^{\dagger})}
U_{i_1j_1}\dots U^{\dagger}_{k_1l_1}\dots\over\int d\phi[dU]e^{-{\rm
Tr}(\sum V[\phi]-\sum \phi U\phi U^{\dagger})}}
\label{correl}
\endequation
$$\approx{\int d\bar\phi[dU]e^{{\rm Tr}(\sum
\bar\phi U\bar\phi U^{\dagger})}
U_{i_1j_1}\dots U^{\dagger}_{k_1l_1}\dots\over\int
d\bar\phi[dU]e^{{\rm Tr}(\sum 	\bar\phi U\bar\phi U^{\dagger})}} $$
In this integral, the scalar field is written at $\phi=V\phi_D
V^{\dagger}$ with $\phi_D$ a diagonal matrix.  The eigenvalues of
$\phi$ are fixed at the value of the master field $\bar\phi$ and, in
order to obtain gauge invariance of the correlator, the angular
matrices $V$ are still integrated,
\equation
d\bar\phi\equiv
\Delta(\bar\phi)DV
\endequation
In any gauge invariant correlator, the matrices $V$ can be absorbed by
redefining $U$.

If we consider for the moment surfaces which are not self-intersecting
so that the filled Wilson loop correlator has at most one
$UU^{\dagger}$ pair on any link we need to consider only the two field
correlator $<U_{ij}U_{kl}^{\dagger} >$. Gauge invariance implies
that\cite{kmsw1}
\equation
<U_{ij}U_{kl}^{\dagger} >=C_{ij}\delta_{il}\delta_{jk}~~~{\rm with}
{}~~~C_{ij}={\int d\bar\phi[dU]e^{N\sum{\rm Tr}\left( \bar\phi U\bar\phi
U^{\dagger}\right)} \vert U_{ij}\vert^2\over\int d\bar\phi
[dU]e^{N\sum{\rm Tr}\left( \bar\phi U\bar\phi U^{\dagger}\right)}}
\endequation
Thus, in the master field approximation, the expectation value of the
filled Wilson loop is given by
\equation
<W_F[\Gamma]>=\sum_S\mu(S) \left[\left(\prod_{\buildrel{ {\rm
sites}}\over{x\in S}} \sum_{i(x)=1}^N\right)\prod_{\buildrel {\rm
links}\over{<xy>\in S}}C_{i(x)~j(x)}\right]
\label{fwl}
\endequation
This is a generalized Potts model on a random surface in which
$N$--component spins reside at each site and the Boltzmann weights
$C_{ij}$ for the bonds are correctly normalized to be conditional
probabilities; $\sum_i C_{ij}$=$1$, $\sum_j C_{ij}$=$1$.

\section{\bf COMPUTING CORRELATION FUNCTIONS}

Techniques for evaluating $C_{ij}$ for general $N$ and for arbitrary
$\bar\phi$ are presented in \cite{shat} and \cite{moroz}. An explicit
formula for $SU(2)$ is given in \cite{kmsw1}. Although the general
formula for $C_{ij}$ in $SU(N)$ is quite difficult to deal with, it is
still possible to estimate the surface dependence of the statistical
model partition function in (\ref{fwl}) when the master field
$\bar\phi$ is homogeneous by considering two different limits.

\subsection{Low Temperature}

First, consider the case where $\bphi_i$ are large.  We also assume
that the eigenvalues $\bphi_i$ are not too close to each other in the
sense that
\equation
\sum_{i\neq j}
{1\over{\left(\bphi_i-\bphi_j\right)^2}} << N
\endequation
(Note that this is a single sum over $j$ for fixed $i$.).  The
integral in (\ref{itzub}) is known to be exact in the semi-classical
approximation (see\cite{kmsw1} for a discussion).  The classical
equation of motion is
\equation
\bigl[ U\bphi U^{\dagger},
\bphi\bigr]=0
\endequation
which, since $\bphi$ is diagonal, is solved by any $U$ of the form
$U_0=DP$ where $D$ is a diagonal unitary matrix and $P$ is a matrix
which permutes the eigenvalues,
\equation
(P\bphi P)_{ij}=\delta_{ij}
\bphi_{P(i)}
\endequation
Also, when N is large and $\bphi$ is not too small the identity
permutation gives the smallest contribution to the action in
(\ref{itzub}) and therefore is the dominant classical solution.  In
this case we use this minimum to evaluate the correlators,
\equation
	I_{IZ}^{-1} \int[dU]e^{N\sum\bar\phi_i\bar\phi_j\vert U_{ij}
\vert^2}U_{i_1j_1}\ldots 	U_{i_nj_n}U_{k_1l_1}^\ad\ldots
U_{k_nl_n}^\ad 	 = \delta_{i_1j_1}\ldots
\delta_{k_nl_n}S^{i_1\cdots i_n}_{k_1\dots k_n} 	\lab{cor}
\endequation
where we have written the normalized integral over diagonal matrices
\begin{eqnarray}
	&S^{i_1\cdots i_n}_{k_1\dots k_n}=\int\prod_\ell
d\theta_\ell \prod_{p<q}
\sin^2(\theta_p-\theta_q)e^{i\left(\theta_{i_1}+\dots+ 	\theta_{i_n}-
\theta_{k_1}-\dots\theta_{k_n}\right)}/\int\prod_\ell
d\theta_\ell\prod_{p<q} 	\sin^2(\theta_p-\theta_q) &
\nonumber\\
\nonumber\\
	& =\left\{\matrix{ 1&{\rm~if~} i_1\dots i_n{\rm ~is ~a
{}~permutation~ 	of~} k_1\dots k_n\cr 0&{\rm~otherwise}\cr}\right.%
\end{eqnarray}
We have decomposed the integration over unitary matrices into an
integration over the diagonals and an integration over the unitary
group modulo diagonals\cite{footn1}. The diagonals are the `zero
modes' for the semiclassical integral and must be integrated exactly.
The unitary modulo diagonal integral is damped by the integrand and is
performed by substituting the classical configuration.  Of course, to
get the next to leading order the latter integration must be done in a
Gaussian approximation.  It can be done for the first few correlators,
The result is $$ 	 C_{ii}=1-{1\over N} \sum_{k\neq i} {1\over
{\left(\bphi_i-\bphi_k\right)^2}} ~~~{\rm when~}i\neq
j~~~C_{ij}={1\over N} {1\over{\left( 	\bphi_i-\bphi_j\right)^2}} $$

We remark that similar calculations can be easily done for correlators
of more than two $U$'s
\equation
	C_{ij,kl}=I_{IZ}^{-1} \int [dU] e^{N\sum \bar\phi_i\bar\phi_j
\vert U_{ij}\vert^2} |U_{ij}|^2 |U_{k \ell}|^2
\endequation
$$ 	i=j~~,~~k=\ell~~~~ C_{ii,kk}=1-{1\over N} \sum_{n\neq i}
\left[ 	\frac{1}{(\bphi_i-\bphi_n)^2}+\frac{1}{(\bphi_k-\bphi_n)^2}
\right] $$
\equation
	i\neq j~~,~~k=\ell~~~~~~~~~~~~~~~~~ 	~~~~ C_{ij,kk}={1\over
N} \frac{1} 	{(\bphi_i-\bphi_j)^2}~~~~~~~~~~~~~~~
\endequation
$$ 	i\neq j~~,~~k\neq \ell~~~~ C_{ij,kl}={1\over N^2} \frac{1}
{(\bphi_i-\bphi_j)^2} \frac{1}{(\bphi_k-\bphi_\ell)^2} 	\left(
1+\de_{ik}\de_{j\ell} 	\right) ~, $$ where the next corrections will
be of the form
\equation
{1\over
N^2}{1\over{\left(\bphi_i-\bphi_j\right)^2}} \sum_{m\neq k}
{1\over{\left(\bphi_k-\bphi_m\right)^2}}~.
\endequation

We call this limit of large $\bar\phi$ the ``Low Temperature'' limit
since in this limit
\equation
	C_{ij}^{\mbox{\tiny\bf LT}}=\delta_{ij}+\dots
\endequation
and the value of the spin at each site is equal. In this case the
$C_{ij}$ represent the Bolzman weight for a perfectly ordered system.
These two cases lead to profoundly different behavior for the filled
Wilson loop.  We shall assume that, by choosing the potential for the
scalar field in (\ref{phii}) appropriately, either of these limits
could be obtained (the eigenvalue repulsion due to the Vandermonde
determinants in (\ref{phii}) and the possibility of adding repulsive
central potentials makes the low temperature limit more natural).

\subsection{High Temperature}

The other limit is where $\bar\phi$ is small.  There, we can obtain
the correlators by Taylor expansion,
\equation
C_{ij}=\int [dU]\left(1+N{\rm Tr} \bar\phi U\bar\phi
U^{\dagger}+\dots\right) \vert
U_{ij}\vert^2=\frac{1}{N}+\frac{\bar\phi_i\bar\phi_j}{N}+\ldots
\endequation

We call this the ``High Temperature'' limit since in this limit is
independent of $i$ and $j$. It thus represents the Bolzman weights for
a highly disordered system.

\subsection{Renormalization of the String Tension}

We begin by estimating the value of the filled Wilson loop for a fixed
surface $S$.  In the ``high temperature'' case the statistical model
is disordered.  The sums over configurations at the various sites are
independent and they contribute an overall factor $N^{V}$ (where $V$
is the number of vertices on the surface) to the expectation value of
the filled Wilson loop.  Furthermore each link contributes a factor
$C_{ij}$$=$$1/N$ so that the links contribute a total factor of
$N^{-L}$ where $L$ is the total number of links. It follows that the
expectation value of the filled Wilson loop goes like
\equation
	<W_F[\Gamma,S]>^{\bf HT}\sim N^{V-L} = N^{2-2g(S)} N^{-A(S)}
\endequation
 where $A(S)$ is the area and $g(S)$ is the genus of the surface $S$
(i.e. the number of plaquettes comprising $S$) and we have used
Euler's theorem, $\chi \equiv 2-2g = V -L + A$.  We thus get the
renormalization of the string tension $\delta\alpha_{\bf HT} =\log N$.
Notice also that higher genus surfaces are suppressed and that the
loop (genus) expansion parameter is $1/N^{2}$.  This is precisely what
is obtained in the conventional strong coupling expansion of Wilson's
lattice gauge theory which is known to describe a string theory with
extra degrees of freedom associated with self-intersections of the
string
\cite{kazak}.

In the ``low temperature'' case, the statistical system is ordered.
The spins on all the sites are frozen at a uniform value.  In this
case the partition function is proportional to the degeneracy of the
ground state,
\equation
	<W_F[\Gamma,S]>^{\mbox{\tiny\bf LT}}=N
\endequation
Note that in this case the statistical model gives no contribution to
the string tension ($\delta\alpha_{\mbox{\tiny\bf LT}}
\approx 0$) and there is no suppression of higher genus surfaces.

In order to proceed to the evaluation of the filled Wilson loop we
need to choose a weight function $\mu(S)$ in order to perform the sum
over surfaces. The most reasonable criterion for choosing such a
weight function is our desire to get a finite physical string tension
in the continuum limit. In order to accomplish this goal we must
choose a weight function $\mu(S)$ which depend on the area of the
surface differently in the low and in the high temperature cases.  It
is known that the number of closed surfaces with a given area grows
exponentially as
\equation
	n(A)\sim A^{\kappa(g)}e^{\mu_0 A}
\endequation
where $\kappa(g)$ is a universal constant which depends only on the
genus of the surface and $\mu_0$ is a non-universal, regulator
dependent constant\cite{zam} which will lead to a renormalization of
the string tension.  In our case, although the surfaces are open, the
above formula should still be valid for surfaces whose area is much
larger than the area of the minimal surface bounded by $\Gamma$.  If
the continuum limit of our theory is realized in the ``high
temperature'' phase we should use the weight function
$\mu_{\mbox{\tiny\bf HT}}(S)\sim N^{A(S)}e^{-\mu_0A(S)}$. This leads
to a vanishing string tension in the lattice theory which is a
necessary condition for having a finite string tension in the
continuum limit.  To accomplish the same goal in the ``low
temperature'' phase we should use $\mu_{\mbox{\tiny\bf LT}}(S)\sim
e^{-\mu_0A(S)}$. Although these choices of $\mu(S)$ give the desired
result, it is rather unnatural to have to choose $\mu(S)$ in such an
ad hoc fashion.

\subsection{Filled Wilson Loops from the Wilson Action}

Fortunately there is a very natural way to obtain the sum over
surfaces in (\ref{fwl}).  Consider the following expectation value
\equation
	<W_F[\Gamma]>={ <W
[\Gamma]e^{\lambda\sum_\Box\left(W[\Box]+W^{\dagger}[\Box]\right)}>
\over <e^{\lambda\sum_{\Box}\left(W(\Box)+W^{\dagger}[\Box]\right)}>}
\label{nfwl}
\endequation
where $W[\Gamma]$ is the conventional Wilson loop. Remember that the
average is weighted by the Kazakov--Migdal action as in
(\ref{correl}): In the master field limit it is computed by
integrating only over $U$--matrices with $\phi=\bar\phi$ and with the
Kazakov--Migdal action. Note that the exponent in (\ref{nfwl}) is
simply the conventional Wilson kinetic term for the gauge fields in
lattice gauge theory. If we expand the right hand side of (\ref{nfwl})
in $\lambda$ the non--vanishing terms are all of those surfaces which
fill the Wilson loop.  The result is thus a filled Wilson loop with a
surface weight $\mu(S)=\lambda^{A(S)}$.

It is clear that we would obtain exactly the same expression (in the
master field approximation) by evaluating the expectation value of the
{\it ordinary} Wilson loop operator in the modified version of the
Kazakov--Migdal model (\ref{qcd}) in which a conventional Wilson term
($\lambda\sum_{\Box}\left(W(\Box)+W^{\dagger}[\Box]\right)$) is added
to the action.  This term breaks the $Z_N$ gauge symmetry explicitly
and allows Wilson loop operators with non--zero area to have non--zero
expectation values. We would expect that it is necessary to keep
$\lambda$ small if one is to maintain the successes of the
Kazakov--Migdal model.  We shall now argue that in the ``low
temperature'' limit this picture is self--consistent in the sense that
the physical string tension is finite when $\lambda$ is small and
consequently the saddle point solution of the original model is
unchanged.  We shall also see that this is not the case in the ``high
temperature'' phase.

Let us begin by determining how $\lambda$ should behave in the
continuum limit if we are to have a finite physical string tension.
As discussed above a necessary condition for having a finite physical
string tension is that the string tension {\it in lattice units}
should vanish. It is thus necessary for the bare string tension
$-\ln\lambda$ to be chosen so as to precisely cancel the
renormalization of the string tension due to both the statistical
model to the sum over surfaces.  It is straightforward to check that
in the ``high temperature'' phase we must choose $\lambda_{\bf
HT}=Ne^{-\mu_0}$, whereas in the ``low temperature'' phase we must
choose$\lambda_{\mbox{\tiny\bf LT}}=e^{-\mu_0}$.  Notice that this
$\lambda$ is proportional to $N$ in the ``high temperature'' phase and
thus cannot be assumed small in large $N$.

In the large $N$ limit of conventional lattice gauge theory the
coefficient of the Wilson term must be proportional to $N$ if one is
to obtain a consistent large $N$ expansion.  In our case we see that
this is true for the ``high temperature'' phase in which case the
Wilson term is of the same order as the Kazakov--Migdal term and it
thus plays an important role in the infinite $N$ limit.  One can say
that in this phase we have ordinary QCD.  Unfortunately it is
impossible to preserve the master field solution of the Kazakov-Migdal
model in this limit since the Wilson term, being of order $N$, would
modify the large N solution, ruining the self-consistency of the
mean-field approximation as described here.

The situation is much more appealing in the ``low temperature'' phase.
In this case the required coefficient of the Wilson term is of order
one. It is subdominant and therefore negligible in the large $N$
limit.  Thus, Migdal's solution \cite{mig} of the Kazakov-Migdal model
in the large $N$ limit should still apply to our proposed modification
of the action.  In fact the only reason that the Wilson term is
important at all in the large $N$ limit of the ``low temperature''
phase is related to the collective phenomenon which orders the
statistical system on the surfaces.  It effectively makes the
statistical model's contribution to the string tension much smaller
than would be expected from naive counting of powers of 1/N and a
truly infinitesimal breaking of the $Z_N$ gauge symmetry
($\lambda_{\mbox{\tiny\bf LT}}/N\rightarrow0$ as $N
\rightarrow\infty$) is sufficient
to make the averages of Wilson loop operators non-vanishing. The
self-consistency of this picture can also be demonstrated by computing
the contribution of the Wilson term to the free energy.  This can be
computed in a small $\lambda$ expansion. For a cubic lattice the
result is:
\equation
Z=<{e}^{\lambda_{\mbox{\tiny\bf LT}}\left(\sum_\Box{\rm
Tr}(W[\Box]+W^{\dagger}[\Box])\right)}> =~Z_{KM}~{\rm exp}{
\left(NV\frac{D(D-1)}{2}\left(\lambda_{\mbox{\tiny\bf
LT}}^2+2\lambda_{\mbox{\tiny\bf
LT}}^6+\ldots\right)\right)}
\label{freeenergy}
\endequation
is of order $N$ (where $V$ is the volume, $D$ is the dimension). This
should be compared with the free energy in the pure Kazakov-Migdal
model which is proportional to $N^2$.  Here, the first term in the
free energy is the contribution of the doubled elementary plaquette
and the second term is due to the two orientations of the elementary
cube.  It is interesting that, to order 6, there is no energy of
interaction of doubled elementary plaquettes with each other. We
conjecture that the interaction energy of surfaces is absent to all
orders and the free energy obtains contributions from all possible
topologically distinct surfaces which can be built from elementary
plaquettes. This suggests a free string picture of the ``low
temperature'' limit of the Kazakov-Migdal model at lattice scales.

\subsection{Self-Intersecting Surfaces}

We have thus far neglected the self-intersecting surfaces in the sum
(\ref{fwl}) which are generated by the expansion of (\ref{nfwl}) in
$\lambda$.  In order to evaluate the contribution of these surfaces we
need to compute the correlator of $n$ $UU^{\dagger}$ pairs on the same
link. The computation of these correlators in full generality is quite
complicated. In the Appendix we compute them in the ``low
temperature'' (ordered) phase.  We find that
\equation
	<U_{i_1j_1}\ldots U_{i_nj_n}U^{\dagger}_{k_1l_1}\ldots
U^{\dagger}_{k_nl_n}> =\delta_{i_1j_1}\ldots\delta_{k_nl_n}S^{i_1\dots
i_n}_{k_1\dots k_n}
\endequation
where $S^{i_1\dots i_n}_{k_1\dots k_n}$ is the tensor which is one if
$i_1\dots i_n$ is a permutation of $k_1\dots k_n$ and is zero
otherwise.  It is now evident that in this limit the $U$--matrices are
replaced by unit matrices which freeze together the spin degrees of
freedom on the various intersecting surfaces.  As a special case we
can consider a single, connected, self--intersecting surface. In this
case all the spin indices on the surface are equal and since
$S^{i_1\dots}_{k_1\dots}=1$ when all arguments are equal the partition
function of the statistical model corresponding to that surface is
simply $N$ just as it was for a non-intersecting surface.  Thus just
as the statistical model does not contribute to the string tension it
also does not contribute to the interaction energy of
self--intersecting surfaces.  This implies that in the ``low
temperature'' limit, the sum over connected surfaces which have a
common boundary behaves like a Nambu-Goto string theory with no
internal degrees of freedom.

\section{\bf DISCUSSION}

In summary, the self-consistency of the ``low temperature'' limit
leads us to a new large N limit of the conventional lattice gauge
theory coupled to scalars: $$ 	Z=\int d\phi
[dU]\exp\left(-N\sum_x{\rm Tr}V[\phi(x)]+N\sum_{<x,y>}{\rm
Tr}\phi(x)U(xy)\phi(y)U^{\dagger}(xy)+\right.  $$
\equation
	\left.+ 	\lambda\sum_{\Box}
(W(\Box)+W^{\dagger}(\Box))\right) 	\label{nkmm}
\endequation
The conventional large $N$ limit occurs when $\lambda $ is of order
$N$ and describes scalar QCD.  The other limit occurs when
$N\rightarrow \infty$ with $\lambda$ of order one. This model is
soluble using the Kazakov--Migdal approach.

It is the latter case in which $\lambda$ remains constant that is of
special interest to us.  In this case we saw that the large $N$
expansion corresponds to a string theory with some unusual features.
The partition function and the Wilson loop expectation value can be
described as a sum over surfaces.  What is unusual is that the genus
of the surfaces is not suppressed in the large $N$ limit, as it is in
continuum QCD.  (We do of course expect the higher corrections in
$1/N$ to suppress higher genus terms.) For a continuum string theory
this sum over the genus is badly divergent.  This, together with the
presence of tachyons, suggests that the true ground state of the
string theory is some sort of condensate.  This could pose a
complication for the present version of the Kazakov-Migdal model in
the continuum limit and deserves further attention.  It is still a
mystery to us how the sum over all surfaces at the lattice scale
should turn into the sum over planar diagrams in the continuum theory
of QCD.

An alternative to the model presented here is the mixed model which
was invented by Migdal \cite{mixed} to solve the problem of $Z_N$
symmetry.  It contains heavy quarks in the fundamental representation
of the gauge group. Despite the obvious differences between our model
and Migdal's mixed model they have many features in common.  As in all
cases when there are fields in the fundamental representation, the
asymptotics of the Wilson loops in the mixed model exhibit a perimeter
law.  In conventional QCD one would expect that if the quarks are
heavy enough, there is an area law for small enough loops, i.e. there
would exist a size scale which is far enough into the infrared region
that the quark potential is linear but the interaction energy is not
yet large enough that it is screened by producing quark-antiquark
pairs.  Thus, in QCD we expect that adding heavy quarks would not ruin
the area law for Wilson loops smaller then some scale.

The mixed model has just the opposite scenario, it is possible to get
an area law only when the heavy quarks are light enough.  This is a
result of the fact that, in the Kazakov-Migdal model, no Wilson loops
are allowed at all unless the $Z_N$ symmetry is explicitly broken.  In
the mixed model, the $Z_N$ charge of links in a Wilson loop must be
screened by the heavy quarks.  This can happen in two ways.  First,
the Wilson loop can just bind a heavy quark to form an adjoint loop -
giving a perimeter law for the free energy of the loop.  This is the
leading behavior if the fermion mass, $M$, is large. The free energy
would go like $1/M^P$ where $P$ is the perimeter.  The only way an
area law might arise is when the fermions are light enough that their
propagators could from a filled Wilson loop with free energy
$1/M^{2L}$ where $L\approx 2A$ is the number of links.  Then, since
the entropy for filled loops is much larger than that for adjoint
loops, these configurations would be important if $M^4<e^{\mu_0}$.
Then, the asymptotics behavior of the Wilson loop would still have a
perimeter law but there would be loops with $4A-P<\mu_0/\ln M$ where
there would be an approximate area law.

{\bf Acknowledgement} This work is supported in part by the Natural
Sciences and Engineering Research Council of Canada and by the
National Science Foundation grant \# NSF PHY90-21984.


\vspace{13cm}

\begin{thebibliography}{99}



\bibitem{dksw}M. Dobrolyubov, I. Kogan, G. Semenoff and N. Weiss,
``Induced QCD Without Local Confinement'', Phys. Lett., in press,
1993.
\bibitem{lgn}G. t'Hooft, Nucl. Phys. B72, 461 (1974).
\bibitem{kmm}
V.A.Kazakov and A.A.Migdal, {\it Induced QCD at large N}, Paris /
Princeton preprint LPTENS-92/15 / PUPT-1322, (June, 1992)
\bibitem{mig}
A.A.Migdal, {\it Exact solution of induced lattice gauge theory at
large N}, Princeton preprint PUPT-1323 (June, 1992)
\bibitem{mig2}
A.A.Migdal, {\it 1/N expansion and particle spectrum in induced QCD},
Princeton preprint PUPT-1332 (July, 1992)
\bibitem{ksw}
I.I.Kogan, G.W.Semenoff and N.Weiss, {\it Induced QCD and hidden local
$Z_N$ symmetry}, UBC preprint UBCTP-92-022 (June, 1992), Phys. Rev.
Lett., in press.
\bibitem{kmsw1}
I.I.Kogan, A.Morozov, G.W.Semenoff and N.Weiss, {\it Area law and
continuum limit in "induced QCD"}, UBC preprint UBCTP-92-026 (July,
1992), Nucl. Phys. B, in press.
\bibitem{GS92}
A.Gocksch and Y.Shen, {\it The phase diagram of the $N=2$
Kazakov-Migdal model}, BNL preprint (July, 1992)
\bibitem{KhM92}
S.B.Khokhlachev and Yu.M.Makeenko,{\it The problem of large-N phase
transition in Kazakov-Migdal model of induced QCD }, ITEP-YM-5-92,
(July, 1992)
\bibitem{CAP92}
M.Caselle, A.D.'Adda and S.Panzeri, {\it Exact solution of D=1
Kazakov-Migdal induced gauge theory}, Turin preprint DFTT 38/92 (July,
1992)
\bibitem{Gr92}
D.J. Gross, {\it Some remarks about induced QCD}, Princeton preprint
PUPT-1335 (August, 1992)
\bibitem{M92}
Yu. M. Makeenko, {\it Large-N reduction, Master field and loop
equations in Kazakov-Migdal model}, ITEP-YM-6-92 (August, 1992)
\bibitem{KM92}S. Khokhlachev and Yu. Makeenko, {\it Adjoint Fermions
Induce QCD}, ITEP-YM-7-92 (August, 1992).
\bibitem{kmsw2}
I.I.~Kogan, A.Morozov, G.W.~Semenoff and N.~Weiss, {\it Continuum
limits of "induced QCD": lessons of the Gaussian model at $ D=1 $ and
beyond}, UBCTP 92-27, ITEP-M7/92 (August, 1992), Int. J. Mod. Phys.,
in press.
\bibitem{mixed}
A.A.Migdal, {\it Mixed model of induced QCD}, Paris preprint
LPTENS-92/23, (August, 1992).
\bibitem{mig100}A. Migdal, {\it Phase Transitions in Induced QCD},
LPTENS-92/22 (August, 1992).
\bibitem{shat}
S.L. Shatashvili, {\it Correlation functions in the Itzykson-Zuber
model}, IAS preprint IASSNS-HEP-92/61, (September, 1992).
\bibitem{moroz}
A. Yu. Morozov, {\it Correlators in the Itzyksen-Zuber integral},
ITEP-M10/92, (September, 1992).
\bibitem{boulatov}D. V. Boulatov,{\it Local Symmetry in the
Kazakov-Migdal Model}, NBI-HE-92-62 (September, 1992).
\bibitem{russakov}B. Russakov, {\it From Hermitean matrix model to
lattice gauge theory}, TAUP 1996-92 (September, 1992).
\bibitem{mig500}A.Migdal, {\it Bose condensation and $Z_N$ symmetry breaking
in the mixed model of induced QCD}, PUPT-1343 (September, 1992).
\bibitem{boulatov2}D. V. Boulatov, {\it Infinite-tension strings at
d$>$1}, NBI-HE-92-78 (November, 1992).
\bibitem{kazak}V. A. Kazakov, Phys. Lett. 128B, 316 (1983);
Zh. Eksp. Teor. Fiz. 85, 1887 (1983) [Sov. Phys. JETP 58 1096 (1983).
\bibitem{zam} A. B. Zamolodchikov, Phys. Lett. 117B, 87 (1982).
\bibitem{itzub}
C.Itzykson and J.B. Zuber, J. Math. Phys. {\bf 21} (1980), 411.
\bibitem{harish}
Harish-Chandra, Amer. J. Math. {\bf 79} (1957) 87.
\bibitem{footn1}Despite the simple form of $S$ its tensor
structure is quite complicated. For example $$
S^{i_1i_2}_{k_1k_2}=\delta^{i_1}_{k_1}\delta^{i_2}_{k_2}
+\delta^{i_1}_{k_2}\delta^{i_2}_{k_1}
-\delta^{i_1}_{i_2}\delta^{k_1}_{k_2} $$
\end{thebibliography}
\end{document}